\documentclass[aps, twocolumn, a4paper, superscriptaddress]{revtex4}

\usepackage{graphicx}
\usepackage{amsmath, amssymb}
\usepackage{lmodern}
\usepackage[T1]{fontenc}
\usepackage[utf8]{inputenc}
\usepackage[english]{babel}
\usepackage{float}
\usepackage{color}

\begin{document}

\title{How directional mobility affects coexistence in rock-paper-scissors models}

\author{P.P. Avelino}
\email[Electronic address: ]{Pedro.Avelino@astro.up.pt} 
\affiliation{Instituto de Astrof\'{\i}sica e Ci\^encias do Espa{\c c}o, Universidade do Porto, CAUP, Rua das Estrelas, PT4150-762 Porto, Portugal}
\affiliation{Departamento de F\'{\i}sica e Astronomia, Faculdade de Ci\^encias, Universidade do Porto, Rua do Campo Alegre 687, PT4169-007 Porto, Portugal}
\author{D. Bazeia}
\email[Electronic address: ]{bazeia@fisica.ufpb.br} 
\affiliation{Departamento de F\'{\i}sica, Universidade Federal da
Para\'{\i}ba 58051-900 Jo\~ao Pessoa, PB, Brazil}
\author{L. Losano}
\email[Electronic address: ]{losano@fisica.ufpb.br} 
\affiliation{Departamento de F\'{\i}sica, Universidade Federal da
Para\'{\i}ba 58051-900 Jo\~ao Pessoa, PB, Brazil}
\author{J. Menezes}  
\email[Electronic address: ]{jmenezes@ect.ufrn.br} 
\affiliation{Instituto de Astrof\'{\i}sica e Ci\^encias do Espa{\c c}o, Universidade do Porto, CAUP, Rua das Estrelas, PT4150-762 Porto, Portugal}
\affiliation{Escola de Ci\^encias e Tecnologia, Universidade Federal do Rio Grande do Norte\\
Caixa Postal 1524, 59072-970, Natal, RN, Brazil}
\affiliation{Institute for Biodiversity and Ecosystem Dynamics, University of Amsterdam, Science Park 904, 1098 XH Amsterdam, The Netherlands}
\author{B.F. de Oliveira}  
\email[Electronic address: ]{breno@dfi.uem.br} 
\affiliation{Departamento de F\'{\i}sica, Universidade Estadual de
Maring\'a, Av. Colombo 5790, 87020-900 Maring\'a, PR, Brazil}
\author{M.A. Santos}  
\email[Electronic address: ]{marcioanicete@gmail.com} 
\affiliation{Departamento de F\'{\i}sica, Universidade Estadual de
Maring\'a, Av. Colombo 5790, 87020-900 Maring\'a, PR, Brazil}

\begin{abstract}
This work deals with a system of three distinct species that changes in time under the presence of mobility, selection, and reproduction, as in the popular rock-paper-scissors game. The novelty of the current study is the modification of the mobility rule to the case of directional mobility, in which the species move following the direction associated to a larger (averaged) number density of selection targets in the surrounding neighborhood. Directional mobility can be used to simulate eyes that see or a nose that smells, and we show how it may contribute to reduce the probability of coexistence.	
\end{abstract}

\maketitle

\section{INTRODUCTION}

It is a known fact in the study of population dynamics that mobility affects species coexistence. As
pointed out by several authors \cite{Reichenbach-N-448-1046,
Frey-PA-389-4265, He-PRE-82-051909, He-EPJB-82-97,
Jiang-PRE-84-021912, Cheng-SR-4-7486, Szczesny-PRE-90-032704,
Szolnoki-JRSI-11-0735}, mobility may contribute to the extinction of some species in systems of several
species that interact cyclically. In particular, in the recent
work \cite{Szolnoki-JRSI-11-0735} the authors review several aspects
of the cyclic dominance in evolutionary games, including pattern
formation and the impact of mobility that motivates the current study.

In the tritrophic systems modeled by the rock-paper-scissors game, coexistence is reached only if all species persist. This model is widely applied to describe biological systems composed of three cyclic, non-hierarchical interacting species, like strains of colicinogenic Escherichia coli \cite{Coli}.

Mobility plays a crucial role in promoting or destroying coexistence of species in rock-paper-scissors games \cite{Reichenbach-N-448-1046}. The movement of individuals can be motivated by the geographic distribution of competitors, leading to directional dispersal of individuals on the grid. For example, in Refs.~\cite{Vicsek-PRL-75-1226, Cambui-IJMPB-28-1450094,Cambui-PA-444-582}, the authors have studied particles moving in the average direction of the particles in their neighborhood. Furthermore, individuals can move following the fluid in which the species are dispersed \cite{Karolyi-JTB-236-12,Yang-C-20-2}. 

In recent works \cite{Avelino-PRE-86-031119,
Avelino-PRE-86-036112, Avelino-PLA-378-393, Avelino-PRE-89-042710}, we
have studied dynamical, geometrical and topological properties of competing networks that
depend crucially on the mobility, reproduction, and competition interactions, assuming a standard mobility.
In the current work, we focus on how directional mobility modifies
the dynamics of the system, with a particular focus on its impact on species coexistence. The motivation is to
make mobility more realistic, by introducing taxis, which allows individuals to walk towards a specific direction defined by an external (local) stimulus \cite{Motivation1}. This behavioral response is a characteristic of various species, e.g.,  rotifers are sensitive to predation risk, and move towards conspecifics and thus diffuse less at higher densities \cite{rotifiers,Motivation2, Motivation3}. Here, we aim to allow individuals to choose the direction to move, based on the spatial distribution of selection targets in the neighborhood.

To make the investigation easier to follow, we consider the system described by three distinct species,
$A$, $B$, and $C$, and in Sec.~\ref{sec2} we describe how the stochastic rules are implemented in our spatial system defined
in a square lattice, and we also explain how the directional mobility is modeled. 
In Sec.~\ref{sec3},  we deal with the time evolution of the system, firstly reproducing the typical spiral patterns that appear in the standard situation and then studying the modifications associated with the introduction of directional mobility. We then investigate temporal features of the system and the related spatial behavior in the square lattice used to implement the stochastic simulations. 
In Sec.~\ref{sec:mob} we investigate the impact of directional mobility on species coexistence. 
We end the work in Sec.~\ref{sec:end}, with our comments and conclusions.

\section{The Model}
\label{sec2}

In this work, we implement stochastic network simulations in a system
composed of three species that change in time in a square lattice having
$\mathcal{N}$ sites, obeying periodic boundary conditions. 
The initial state is formed by a lattice where each site contains a
single species or is empty. Species and empty sites are initially 
distributed randomly, in such a way that, in the initial state, there are
$\mathcal{N}/4$ sites associated to each one of the three species and the empty
sites.

\begin{figure}[t]
\centering
\includegraphics[scale=1.4]{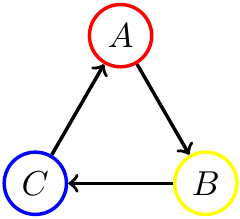}
\caption{The arrows illustrate how selection works in the case of three species that we consider in this paper.}\label{fig1}
\end{figure}

At each time step, an individual is randomly chosen on the grid. This individual can interact with one of its eight immediate neighbors (we are using the Moore vicinity). An interaction can be summarized as follows: select the individual, pick up the neighbor,
choose the rule and implement it in the lattice. The unit of time $\Delta\,t = 1$ is defined as the time necessary for ${\cal N}$ interactions to occur - one generation time. All simulations are done for 15000 generations, with the first 5000 generations discarded, allowing that all the investigation be implemented after the pattern formation.

The stochastic model is characterized by mobility,  selection, 
and reproduction. In the current work, these interactions happen with probabilities $m=0.60$, $r=0.20$, and $p=0.20$, respectively (probabilities are same for all species). Reproduction only happens if a neighbor grid point is empty.  If selection is sorted a random neighbor selection target (if it exists) is substituted by an empty site - it follows the
rock-paper-scissors game, as illustrated in Fig.~1.

\begin{figure}[H]
\centering
\includegraphics[scale=0.64]{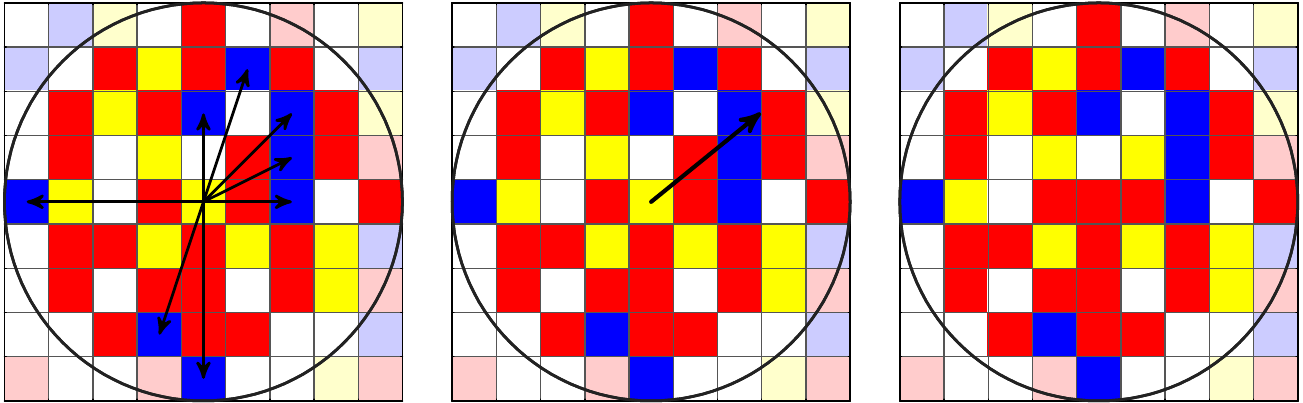}
\caption{As an illustration of the directional mobility, the left panel identifies the possible selection targets of
the central individual inside the euclidean circle of radius $r_c$. The total distance is shown in the middle panel,
and the central individual moves in this direction, as it appears in the right
panel.}\label{fig2}
\end{figure}

Directional mobility is implemented as follows: when mobility is sorted, a region of radius $r_c$ around the chosen individual is delimited. In this region, one vectorially identifies the possible selection targets. The direction of maximum likelihood of finding a the possible selection targets is denoted by ${\vec r}$. The individual will move in the closest direction to ${\vec r}$.
In other words, the individual switches position with its neighbor, in the closest direction to ${\vec r}$. This is illustrated in Fig.~\ref{fig2} for the case $r_c=4$, with $a=1$. Given the cyclicity of the rock-paper-scissors game, directional mobility implies that individuals always prefer running away from hostile regions, by choosing to move towards areas where they dominate.

\begin{figure}
\centering
\includegraphics[scale=0.94]{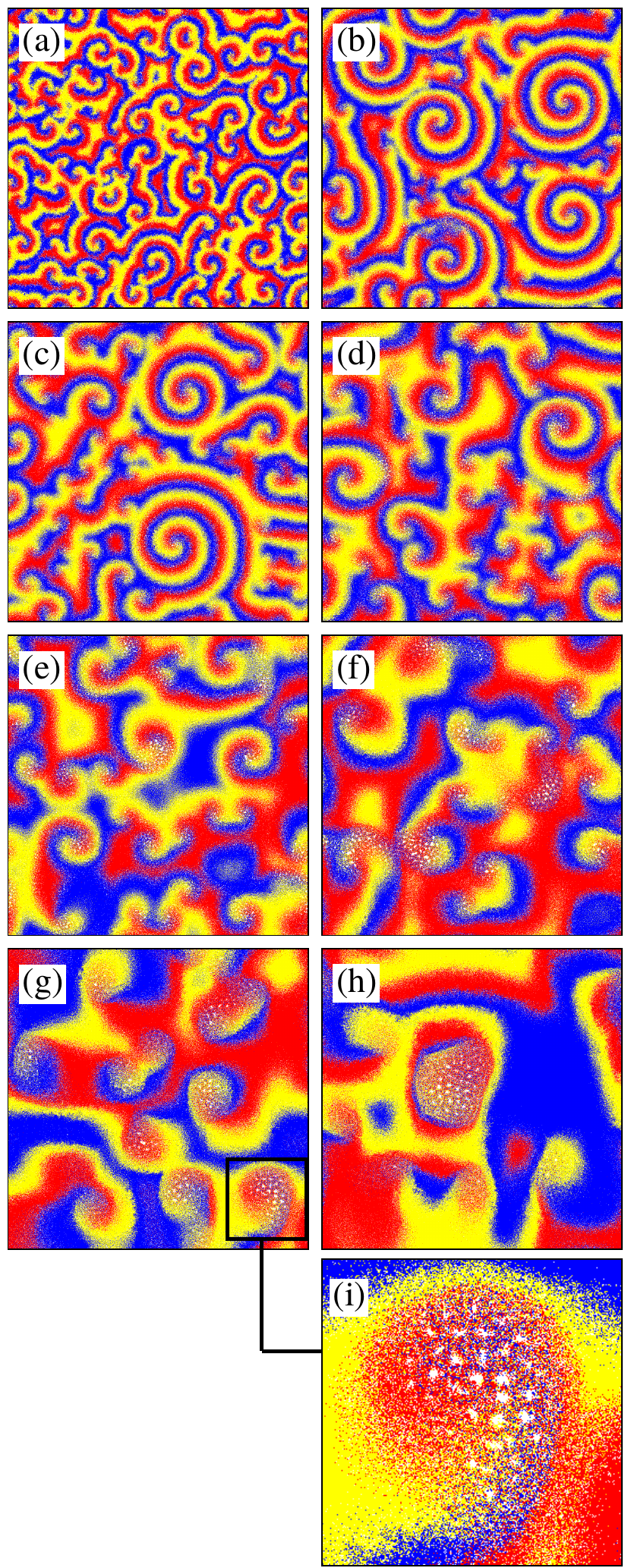}
\caption{Spatial patterns taken from $1000^2$ lattices after $20000$ generations. The panel $(a)$ represents the simulation with standard mobility whereas the panels $(b)$, $(c)$, $(d)$, $(e)$, $(f)$, $(g)$, $(h)$, are snapshots of spatial patterns in simulations with directional mobility for $a=1$, $a=2$, $a=4$, $a=8$, $a=16$, $a=32$, $a=64$, respectively.  Finally, the panel $(i)$ is a zoom of the islands of empty sites present in the selected region of the snapshot of the simulation for $a=32$.}\label{fig3}
\end{figure}

We calculate the vector ${\vec r}$ as
\begin{equation}
{\vec r}=\sum_{i=1}^k d(r_i){\vec r}_i,
\end{equation}
where ${\vec r_i}$ is the distance between the competitors, and $k$ stands for the number of possible selection targets inside the circle of radius $r_c$. Moreover, in this paper we choose the function $d(r_i)$ to be defined by
\begin{equation}
d(r_i)=\exp\left(\displaystyle\frac{-(r_i-1)^2}{a}\right),
\end{equation}
where $a$ is a parameter that controls the maximum reach of the directional mobility. 
In our simulations, we assume $a=2^l$, where $l=0, 1, ..., 6$, and 
cutoff radius $r_c=r_c(a)$ so that $d(r \geq r_c) =0$. Table I shows the cutoff radius and the respective values of $r_c$, used in our simulations. In the case of $r \leq r_c$, we consider $d(r)=0$ for $d(r) < 1 \times 10^{-2}$.

Throughout the paper we will compare the numerical results obtained by assuming directional mobility and the standard case, where individuals move randomly. In the case of standard mobility, individuals are not able to figure out the best direction to move. Therefore, the standard random movement cannot be recovered even if one assumes a small radius of interaction in the directional mobility case.

\begin{table}[h!]
\centering
 \small
\caption[]{The cutoff radius and the respective values of $r_c$.}
\begin{tabular}{lcccccc}
\hline
\hline
\begin{minipage}[t]{.05\textwidth}\begin{flushleft} $a=1$\end{flushleft}\end{minipage}&
\begin{minipage}[t]{.05\textwidth}$2$\end{minipage}&
\begin{minipage}[t]{.06\textwidth}$4$\end{minipage}&
\begin{minipage}[t]{.06\textwidth}$8$\end{minipage}&
\begin{minipage}[t]{.06\textwidth}$16$\end{minipage}&
\begin{minipage}[t]{.06\textwidth}$32$\end{minipage}&
\begin{minipage}[t]{.06\textwidth}$64$\end{minipage}\\
\hline
$r_c\!=\!4$ &$5$	&$6$ &$8$  &$10$&$14$&$19$\\
\hline
\hline
\end{tabular}
\label{tab1}
\end{table}

\section{Results}
\label{sec3}

Up to here, we defined the system and showed how the rules have to be implemented to drive its time evolution and spatial behavior. From now on, we pay closer attention to some of its main features, which we describe below.  

\subsection{Pattern formation}

We first consider the long-time evolution of the system, with standard and directional mobility, controlled by several distinct values of $a$.

Figure ~\ref{fig3} displays snapshots of the patterns obtained for a square lattice of size $1000^2$ after 20000 generations, for the standard mobility, and for various values of $a$. The figure shows that the increase of $a$ changes the spiral patterns, with the formation of clusters of groups of empty sites. This happens because individuals enter domains 
dominated by possible selection targets, moving perpendicular to the interface of empty spaces separating the spiral arms - they move into the direction with a higher density of possible selection targets. As a consequence, islands of empty sites grow along the boundaries between the spiral arms. 

Broadly speaking, the larger $r_c$ the further individuals can move. This means that the increase of the cutoff radius causes the enlargement of the average areas occupied by the domains.  If the area of the domains is the same order (or larger) of the grid size, the probability of extinction of the species increases drastically. As a result, only one species survive since the directional mobility diminishes the chances of species coexistence.

\begin{figure}
\centering
\includegraphics[scale= 1.0]{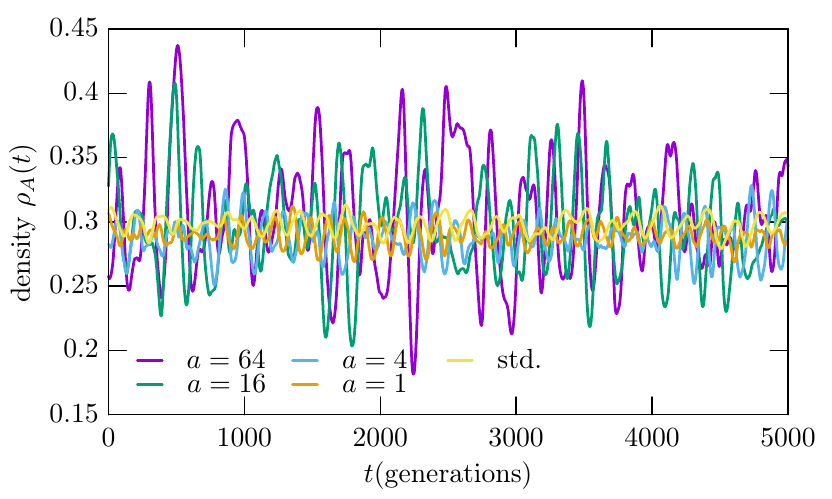}
\caption{Abundance of species $A$ as a function of the time for directional mobility considering various values of $a$. The fluctuation in the size of the population increases as one increases the value of $a$.}
	\label{fig4}
\end{figure}
\begin{figure}
\centering
\includegraphics[scale= 1.0]{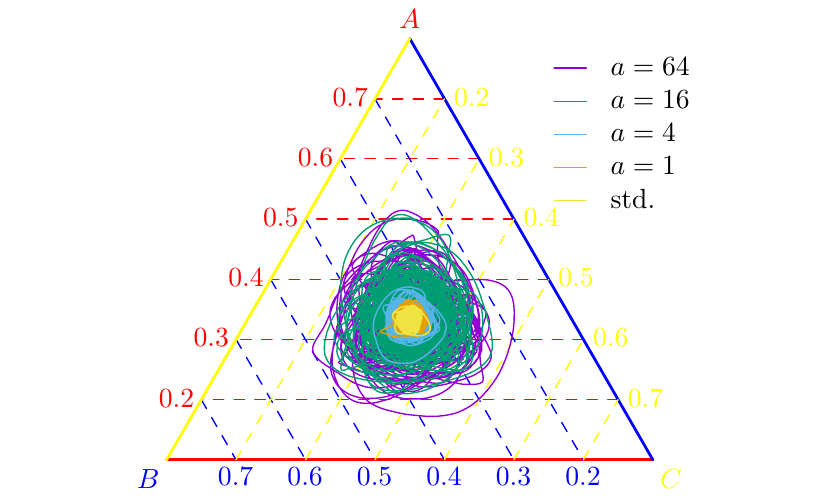}
\caption{This ternary diagram illustrates the evolution of the competition network with three species. The trajectories were taken for a single realization. Although the fluctuations of the population sizes increase with $a$, the species coexist because the average size of the domains is smaller than the grid size. 
}\label{fig5}
\end{figure}

The results displayed in Fig.~\ref{fig3} show that the inclusion of the directional behavior changes the way the species organize themselves in space. They suggest that we investigate both the temporal and the spatial evolution of the species. Hence, below we start focusing on some temporal features of the systems.

\subsection{Temporal behavior}

Let us now investigate some features of the system time evolution. 
Toward this goal, let us first consider how the abundance of a given species changes in time. As all species play a similar role, we will focus on the number density of species $A$ which we denote by $\rho_A(t)$. 

Figure ~\ref{fig4} shows how the abundance of species $A$ changes over time. The data were collected starting counting the time after $10000$ generations. The set of abundances of all species is shown in the ternary diagram in Fig.~\ref{fig5} for several values of $a$.
The results depicted by Figs. \ref{fig4} and \ref{fig5} show that the fluctuation in the size of the population increases as one increases the value of $a$ (or, equivalently, $r_c$). However, no species dies out since the average domain size (that increases with $a$) is smaller than the lattice size. For larger interaction radius, fluctuations may lead to the extinction of species. 

\begin{figure}
\centering
\includegraphics[scale= 1.0]{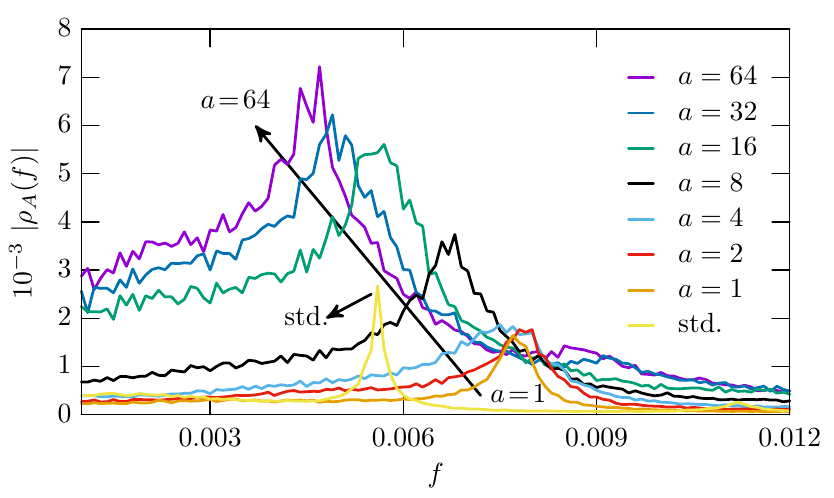}
\caption{The spectral density defined in Eq.~\eqref{dft} is depicted in terms of the frequency, for several distinct mobilities. The spectral density defined in Eq.~\eqref{dft} is depicted in terms of the frequency, for several distinct mobilities. The arrows show how $a$ grows and which curve represents the result provided by simulations with standard mobility.}
\label{fig6}
\end{figure}

In order to further investigate the time evolution of the species, we make a Fourier analysis of $\rho(t)$. Following closely
Refs.~\cite{Washenberger-JPCM-19-065139, He-PRE-82-051909,
He-EPJB-82-97, Cianci-PA-410-66}, we introduce the discrete Fourier transform to get
\begin{equation}
\rho(f) = \frac1{N_G}\,\displaystyle \sum_{t=0}^{N_G-1} \rho(t) \cdot e^{-2\pi i f t}\ ,\label{dft}
\end{equation}
where $f = n/N_G$ with $n = [0, N_G-1]$ and $N_G = 10000$ generations. 
Figure~\ref{fig6} shows the spectral density for the abundance corresponding to species $A$, with the results depicted for an average over 100 simulations with different initial conditions. For increasing values of $a$, the amplitude at maximum frequency $f_{max}$ increases, although $f_{max}$ itself decreases.

\begin{figure}
\centering
\includegraphics[scale= 1.0]{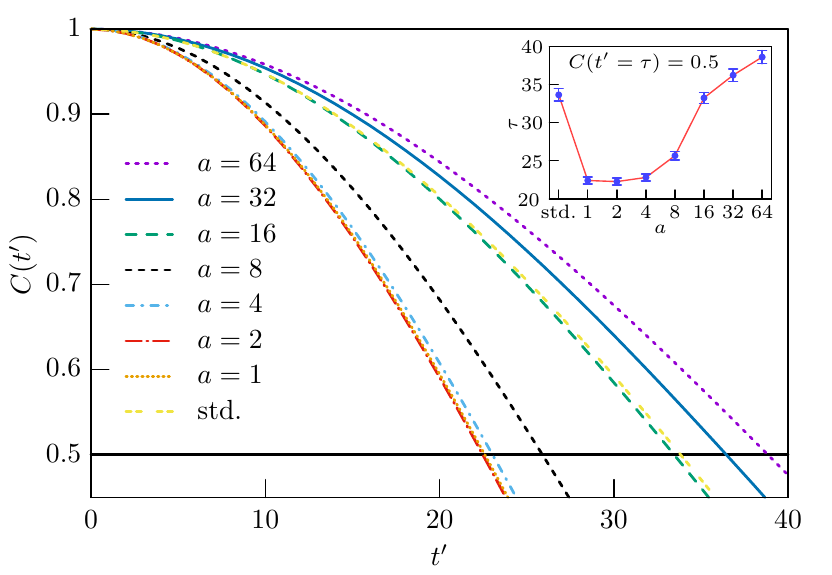}
\caption{Temporal autocorrelation function for various interaction radius. The inset shows the correlation length as a function of $a$.} \label{fig7}
\end{figure}

Another related study concerns the temporal correlation length, $\tau$, which is extracted from the autocorrelation function as $C(t=\tau)=1/2$, that is the time for the autocorrelation to decrease to half of its value at the initial time. Following closely Refs.~\cite{Reichenbach-JTB-254-368, Frey-PA-389-4265, He-PRE-82-051909,
Groselj-PRE-91-033009}, we introduce the autocorrelation function in the form

\begin{equation}
C_{AA}(t')=\frac{1}{C_{AA}(0)}\!\displaystyle\sum^{N_G-t'}_{t=0}\!\!
{\big(\rho_A(t)\!-\!\langle \rho_A \rangle\!\big)\!
\big(\rho_A(t+t')\!-\!\langle \rho_A \rangle\!\big)}\label{act}
\end{equation}
where $\langle\rho_A \rangle$ is the average of the abundance $\rho_A(t)$, for the species $A$. We use Eq.~\eqref{act} to calculate the autocorrelation displayed in Fig.~\ref{fig7} in the case of standard mobility and for several values of $a$. Also, the inset in Fig.~\ref{fig7} shows how the correlation time varies as a function of the parameter $a$. It shows that the correlation time increases as $a$ increases.

\subsection{Spatial behavior}

We now turn attention to the spatial behavior of the system. In order to quantify this behavior we introduce the quantity
\begin{equation}
C(r') = \displaystyle \sum_{|\vec{r}\, '| = x+y} \dfrac{C(\vec{r}\, ')}
{{\rm min}\big(2N-(x+y+1), x+y+1\big)}\ .\label{acr1}
\end{equation}
where $C({\vec r}\, ')$ is the spatial autocorrelation function, calculated from the Fourier transform of the spectral density as
\begin{equation}
C(\vec{r}\, ') =
\dfrac{\mathcal{F}^{-1}\{S(\vec{k})\}}{C(0)}\ .\label{acr2}
\end{equation}
The spectral density $S({\vec k})$ is given by
\begin{equation}
S(\vec{k}) = \displaystyle \sum_{k_x, k_y}
\varphi(\vec{k})\varphi^{*}(\vec{k}) \ ,\label{psd}
\end{equation}
where $\varphi(\vec{k}) = \mathcal{F}\{\phi(\vec{r}) - \langle \phi
\rangle\}$ and $\phi({\vec r})$ represents the species in the position $\vec{r}$ in the lattice; here we are using $0$ for the empty sites, and $1,2,$ and $3$ for species $A, B,$ and $C$, respectively.

\begin{figure}[t]
\centering
\includegraphics[scale= 1.0]{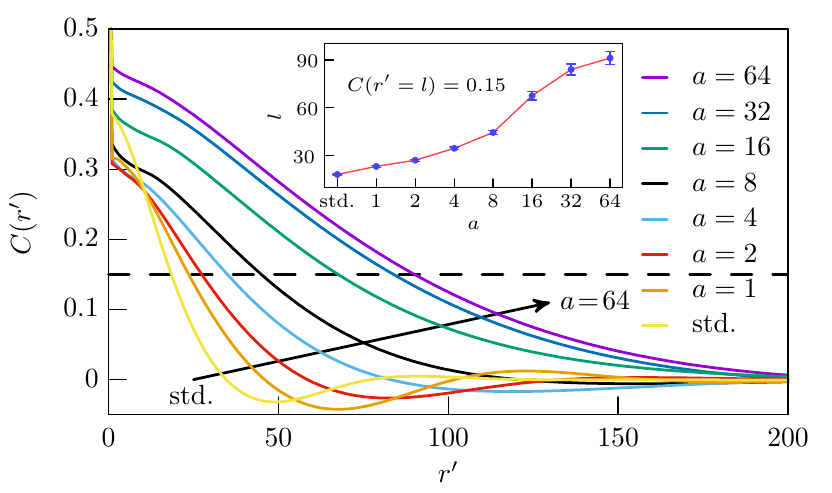}
\caption{Temporal autocorrelation function for various interaction radius. The inset shows the correlation length as a function of $a$. The arrow shows the order of the curves that represent simulations from the standard mobility to directional mobility with $a=32$.}\label{fig8}\end{figure}

\begin{figure}[!htb]
	\centering
		\includegraphics{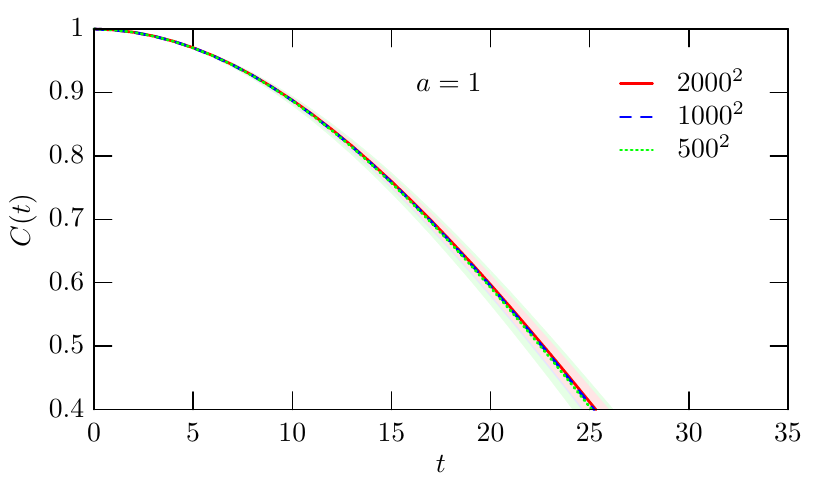}
		\includegraphics{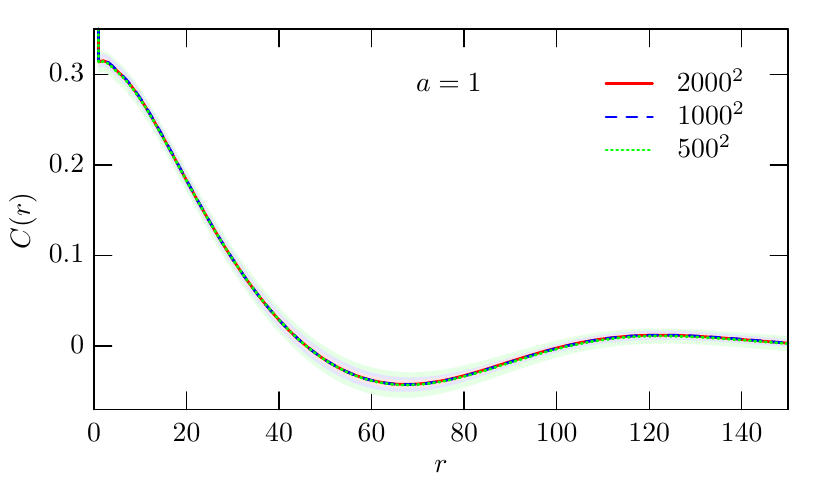}
	\caption{Temporal (upper panel) and spatial (lower panel) autocorrelation functions for different lattice sizes. The margin of error is shown by the shaded area around the functions.}
	\label{fig99}
\end{figure}

Figure ~\ref{fig8} displays $C(r')$ for the directional mobility for various choices of $a$. The inset shows the characteristic length $l$ which we define as $C(r'=l)=0.15$. The results show that the length $l$ increases as one increases the value of $a$. This fact appears clearly in Fig.~\ref{fig3} since there one notes the enlargement of the colored regions which identify the distinct species in the system, as $a$ increases. 
Figure \ref{fig99} shows that both the spatial and the temporal correlation functions do not change significantly 
on the lattice size if $\mathcal{N}/a$ is sufficiently large.

\section{Mobility versus coexistence}
\label{sec:mob}
In the previous sections, we have studied how the systems changes in time and modify their spatial features when one increases the parameter $a$ which controls the directional mobility. Now we will explore further the impact of directional mobility on species coexistence. To this purpose, we run a large number of numerical simulations  with $100^2$ and $200^2$ sites, for $a=1,2,4,8,$ and $32$.
\begin{figure}
\centering
\includegraphics[scale= 1.0]{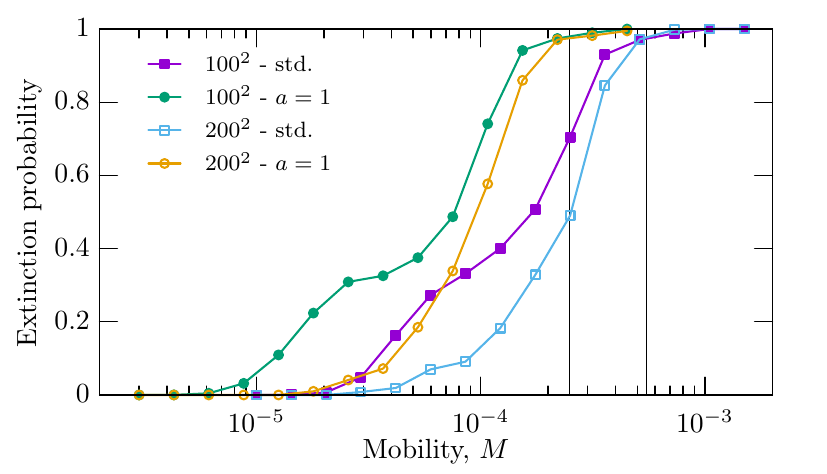}
\caption{The extinction probability as a function of the Mobility M, in the cases of standard mobility and directional mobility with $a=1$, for the lattices with $100^2$ and $200^2$ sites.}
\label{fig9}
\end{figure}
\begin{figure}
\centering
\includegraphics[scale= 1.0]{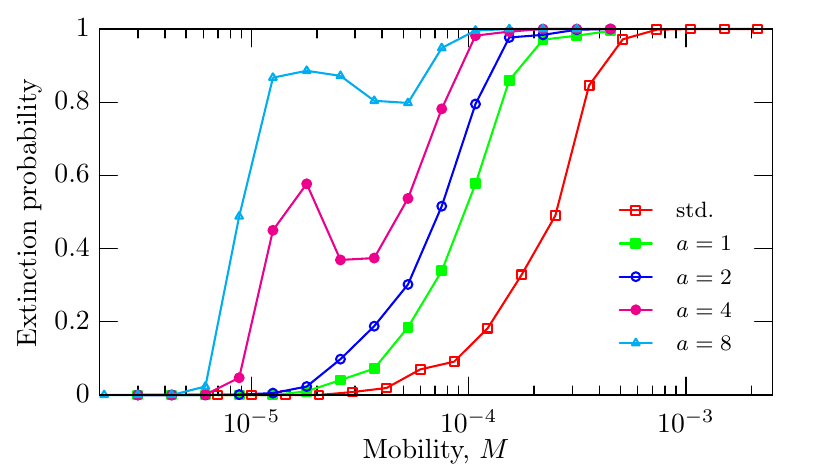}
\caption{The extinction probability as in Fig.~\ref{fig9}, but now for the lattice with $200^2$ sites, and with directional mobility with $a=1,2,4,$ and $8$.}
\label{fig10}
\end{figure}
We introduce some auxiliary parameters $p'$, $r'$ and $m'$ and rewrite the probabilities of selection, reproduction, and mobility as $p'/(p'+r'+m')$, $r'/(p'+r'+m')$ and $m'/(p'+r'+m')$, respectively. 
By setting $p'=r'=1$, the Mobility parameter can be written as  $M=m'/2N^2$, that is proportional to the typical area explored by an individual, per unit time \cite{Reichenbach-N-448-1046}.

Hence we study the extinction probability, that is, the probability of extinction of two species as a function of $M$. The results are shown in Fig.~\ref{fig9}. The vertical lines shows the critical mobilities, $M_c=(5.5\pm0.5)\,10^{-4}$ in the case of standard mobility, and $M_c=(2.5\pm0.5)\,10^{-4}$ for $a=1$. The result for the standard mobility is in good accordance with results obtained in Refs.~\cite{Reichenbach-N-448-1046,Jiang-PRE-84-021912,Cheng-SR-4-7486}, whereas the result for $a=1$ shows that the critical mobility decreases when directional mobility is assumed.

The data of Fig.~\ref{fig9} were taken from the average over 20000 simulations for lattices with $100^2$ sites, and from the average over 1000 simulations for grids with $200^2$ sites. The results displayed in Fig.~\ref{fig9} show that for $a=1$ species go extinct at a smaller value of the critical mobility when compared with the standard case. This happens because the average size of the domains is larger in the case of directional movement, for same mobility probability. 

The results for the lattice with $200^2$ sites, shown in Fig.~\ref{fig10}, confirm those presented in Fig.~\ref{fig9}: directional mobility reduces the probability of coexistence. Moreover, the larger the interaction radius, the more likely the species go extinct.

\section{Comments and conclusions}
\label{sec:end}

In this work, we studied a system of three distinct species that changes in time in cyclic dominance interactions, following the rules of the rock-paper-scissors game. We consider that the movement of the individuals on the grid depends on the spatial distribution of each species. This introduces a directional mobility, that means that individuals move in the direction with a larger number of possible selection targets. As a result,  both the time evolution and the spatial organization of the species change significantly. 
The results show that directional mobility reduces the probability of coexistence. This effect is stronger for larger interaction radius of individuals because the further the individual mobility reaches,  the larger are the average size of the domains are. As long as the average size of the domains is smaller than the grid size, the species coexist, but the increase of the mobility to larger and larger values contributes to the extinction of the species.

The results of the work are of current interest, because directional mobility may contribute to change both the time evolution and the spatial behavior of the system. In particular, one can use directional mobility to model species whose interactions are dependent on space, adding effects due to spatial inhomogeneities in the lattice. These issues open new routes of applications in problems of current interest in several areas of research, including agriculture, ecology, and other related areas of nonlinear science. We hope to report on these and in other related issues in the near future.   

Finally, we point out that there are alternative theories of biodiversity that consider restrictions for the mobility of individuals. For example, in the neutral theory of biodiversity, restricted immigration of organisms from local communities is assumed\cite{Extra1,Extra2,Extra3,Extra4,Extra5}. In this case, a modified version of our stochastic model (constraining how far each individual can reach), can be used to study the effects on the spatial patterns and the population dynamics. We hope to address this issue in future works.

\section*{ACKNOWLEDGMENTS}

We thank Arne Janssen and Maarten \mbox{Boerlijst} for useful discussions. This study was supported by CAPES, CNPq, FAPERN, FCT, Funda\c c\~ao Arauc\'aria, INCT-FCx, and the Netherlands Organisation for Scientific Research (NWO) for financial and computational support. PPA acknowledges support from FCT Grant UID/FIS/04434/2013, DB acknowledges support from Grants CNPq:455931/2014-3 and CNPq:306614/2014-6, LL acknowledges support from Grants CNPq:307111/2013-0 and CNPq:447643/2014-2, and JM acknowledges support from NWO Visitor's Travel Grant 040.11.643. 

\bibliography{ref}
\end{document}